\documentstyle[preprint,aps,epsfig]{revtex}

\tighten
\begin{document}
\draft

\title{Vortex-Line Phase Diagram for Anisotropic Superconductors}    

\author{ W. A. M. Morgado$^{a}$ \footnote{Corresponding author. Fax:+55
21 512-3222. e-mail: welles@fis.puc-rio.br}, M. M. Doria$^{b}$, G.
Carneiro$^{b}$ and I.G. de Oliveira$^{c}$}  

\address{$^a$Departamento de F\'{\i}sica\\Pontif\'{\i}cia Universidade
Cat\'olica  do Rio de Janeiro\\ C.P. 38071 \\ 22452-970, Rio de
Janeiro-RJ, Brazil\\ 
$^b$Instituto de F\'{\i}sica\\Universidade Federal do Rio de Janeiro\\  
C.P. 68528\\ 21945-970, Rio de Janeiro-RJ, Brazil\\
$^c$ Universidade Igua\c{c}u - UNIG, Nova Igua\c{c}u 26260-100, Brazil}
\date{\today}
\maketitle
\begin{abstract}

The zero temperature vortex phase diagram for  uniaxial anisotropic
superconductors placed in an external magnetic field tilted with
respect to the axis of anisotropy is studied for parameters typical
of BSCCO and YBCO. The exact Gibbs free energy in the London
approximation, using a self-energy expression with an anisotropic
core cutoff, is  minimized numerically,
assuming only that the equilibrium vortex state is a
vortex-line-lattice with a single vortex line per primitive unit cell.
The numerical method is based on simulated annealing and uses a fast
convergent series to calculate the energy of interaction between vortex
lines. A phase diagram with three distinct phases is reported and the
phases are characterized in detail. New results for values of the
applied field close to the lower critical field are reported.

\end{abstract}

\narrowtext

\section{Introduction} 
\label{sec.int}

One  important theoretical problem  in the study of type-II
superconductors is  the zero temperature equilibrium vortex phase
diagram for  clean, uniaxially anisotropic, bulk materials subjected to an
external magnetic field, ${\bf H}$ tilted with respect to the axis of
anisotropy ($c$-axis). 	In the case of isotropic materials the vortex
phase is the well known triangular vortex-line lattice with the vortex
lines oriented parallel to ${\bf H}$. In anisotropic materials the
vortex phase is still a vortex-line lattice  but, in general, the lattice is
not triangular nor are the vortex lines  oriented along ${\bf H}$~\cite{rev1,rev2}. 

Previous work find that the vortex phases for the anisotropic 
superconductor  are related to those for the isotropic one when 
${\bf H}$ is oriented along the c-axis, in which case they are identical,
when ${\bf H}$ is along the a-b plane or for $H$ large compared with
the lower critical field~\cite{cdk}. In the latter two cases, the 
vortex-line-lattices are  related to the triangular lattice by scaling
and the vortex lines are essentially parallel to ${\bf H}$. For general
orientations of ${\bf H}$ this is not the case. The vortex phases were first obtained by Buzdin and Simonov~\cite{bs} by numerically  minimizing the Gibbs free-energy in the London limit with material parameters appropriate for YBCO. Our calculation,
like their's, minimizes the London Gibbs free-energy numerically. However, we go beyond the work of these authors by incorporating into our calculations recent results for the vortex state at the lower critical field, not taken into account by them~\cite{sbh,wdg}. We show that this  leads to  new vortex phases close to the lower critical field.  

It was found first by Grishin, Martynovich and Yampol'skii~\cite{gris} and later by others~\cite{vatr1,vatr2,vatr3}, that 
a pair of straight vortex lines parallel to each other and coplanar
with respect to the c-axis  attract at large distances. A
consequence of this is that a chain of such lines 
has lower energy than
the same vortex lines placed far apart. On the basis of 
this result
Buzdin and co-workers~\cite{vatr1,vatr3} suggested that at the lower critical field the
mixed state consists of a vanishing small density of vortex line chains
instead of vortex lines. However this argument does not apply in
general. As pointed out in Ref.\cite{wdg}, in order to determine whether  vortex
lines or vortex-line chains constitute the vortex phase at the lower critical field it is necessary to minimize the Gibbs free-energy for each one of them and obtain
their respective lower critical fields. The one with the lower
critical field is the true equilibrium phase. One important aspect of the
calculation of Ref.\cite{wdg} is the use of an expression for the tilted
vortex-line self energy with  an appropriate anisotropic core cutoff function. This 
expression was derived by Subdo and Brandt~\cite{sb} and later used in Ref.\cite{sbh} to
show, by minimizing the London Gibbs free-energy 
for a vanishing small density of vortex lines, that for anisotropy
parameters typical of BSCCO there is coexistence two such states, with different  vortex lines tilt angles, for one particular orientation of ${\bf H}$. 
These authors also showed that if a self-energy
expression based on an isotropic core cutoff, such as the one adopted
by Buzdin and Simonov~\cite{vatr1}, is used instead, the coexistence disappears. It is 
found in Ref.\cite{wdg} that at the lower critical field 
and for anisotropy parameters typical of BSCCO the equilibrium vortex
state for ${\bf H}$ tilted by $\alpha < \alpha_c=7.9^o$ with respect to
the c-axis is a dilute vortex-line chain state whereas for $\alpha >
\alpha_c=7.9^o$  it is a dilute vortex-line one. For anisotropy
parameters typical of YBCO Ref.\cite{wdg} finds that the lower critical fields
for vortex lines and vortex-line chains differ so little from each other that
it is not possible do decide which one corresponds to true equilibrium.

In this paper, we minimize the exact Gibbs free-energy in the London limit
using  Sudbo and Brandt~\cite{sb} expression the vortex line self energy. We
assume that the equilibrium state the vortex lines are  straight and
parallel to each other and  the VLL has only one vortex line per 
unit cell,  and obtain the vortex lines orientation and VLL shape as
functions of ${\bf H}$,  by minimizing the Gibbs free-energy in the
London limit.

\begin{center} 
\begin{figure}
\epsfig{file=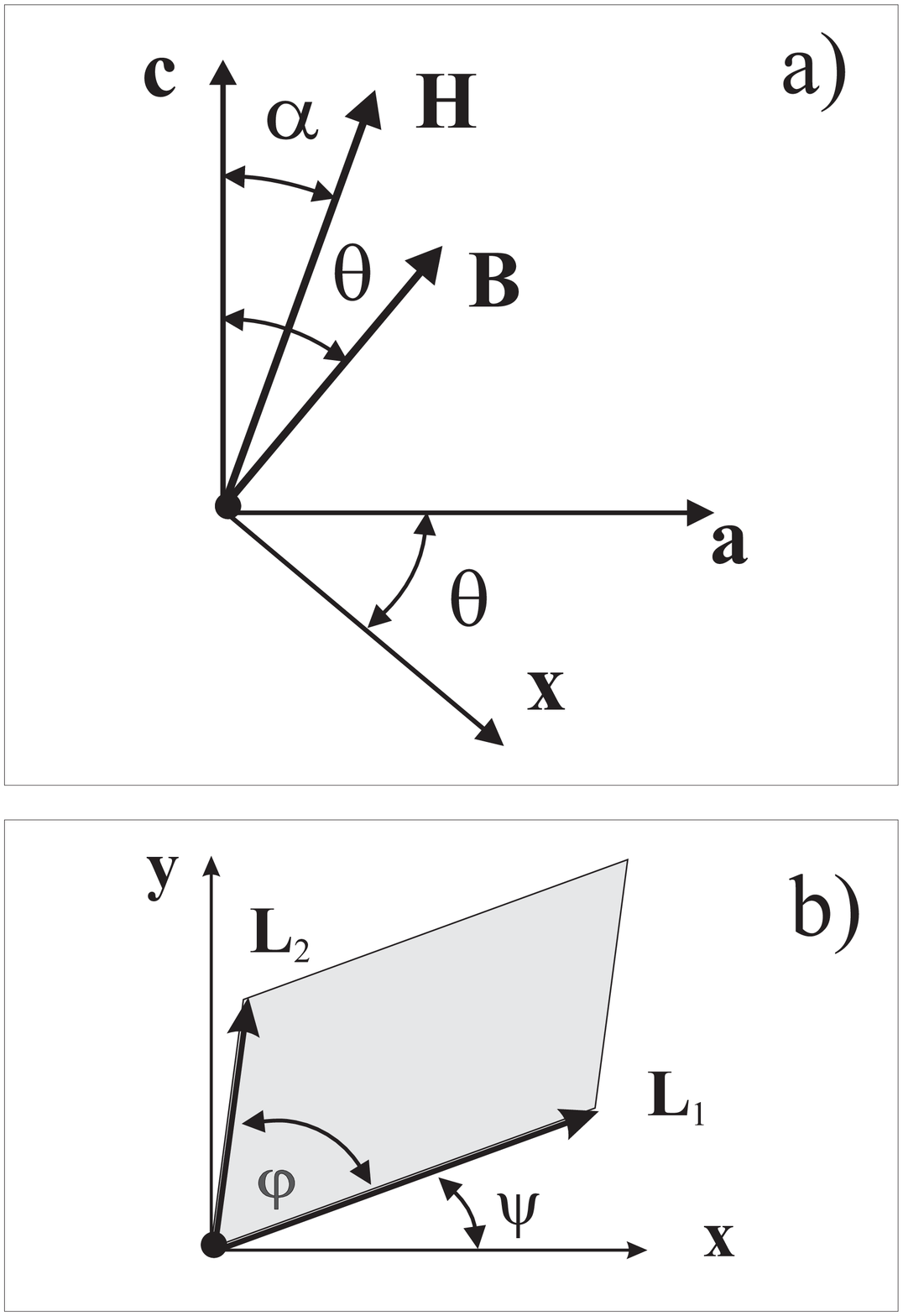,height=8 cm,width=5.65cm,clip=}
\caption{Definition of  coordinate system and VLL primitive unit cell.
a) Relative orientations in the plane containing  ${\bf H}$, the 
$c$-axis and the equilibrium induction ${\bf B}$. The x-axis is
perpendicular to ${\bf B}$ and the $y$-axis is perpendicular to
the plane ${\bf H}$-${\bf B}$, pointing inside. b) Generic VLL
primitive unit cell in the $x-y$ plane.}
\label{fig.hbcu}
\end{figure}
\end{center}

The free parameters in our calculation for  given ${\bf H}$, specified
here by its magnitude, $H$, and its tilt angle with respect 
to the c-axis, $\alpha$, are the
vortex lines tilt angle, $\theta$, and the VLL primitive unit cell vectors
${\bf L_1}$ and ${\bf L_2}$, defined in the plane perpendicular to the
vortex lines. Since the vortex lines are straight and parallel to each
other, the magnetic induction ${\bf B}$  is also tilted by $\theta$
with respect to the c-axis and, as explained in detail in Sec.\
\ref{sec.lt}, it is 
coplanar with ${\bf H}$ and the c-axis. The  orientations of
${\bf H}$, ${\bf B}$, and the c-axis, and the VLL primitive unit cell
are shown in Fig.\ \ref{fig.hbcu}. 
Hereafter we specify the VLL unit cell by the magnitudes of ${\bf L_1}$
and ${\bf L_2}$, denoted by $L_1$ and $L_2$,  by the angle
between ${\bf L_1}$ and ${\bf L_2}$, $\varphi$, and by
and the angle of rotation of the VLL unit cell with respect to
${\bf B}$, $\psi$, as shown in Fig.\ \ref{fig.hbcu}.

As discussed  in Sec.\ \ref{sec.lt},  a minimum of the
Gibbs free-energy  occurs for $\psi=0$, where $\psi$ is defined in
Fig.\ \ref{fig.hbcu}. We 
assume that this is the absolute minimum and carry out the
calculations with  $\psi=0$. This means that ${\bf L_1}$ is also
coplanar with ${\bf H}$ and the c-axis,oriented along the x-direction
defined in Fig.\ \ref{fig.hbcu}. To
simplify the calculations we set  $\varphi=90^o$. This means that the VLL
unit cell is a rectangle, with ${\bf L_2}$ perpendicular to the plane
defined by ${\bf H}$, ${\bf B}$ and the c-axis (Fig.\ \ref{fig.hbcu}).
We find that this simplifications does not change  our main results.
Under these conditions we have only three free parameters, to be determined
by minimizing the Gibbs free-energy: $\theta$, $L_1$ and $L_2$. This
minimization is carried out numerically, using a fast convergent series
to calculate lattice sums entering the expression for the VLL
interaction energy~\cite{mmd}. This series, together with the small number of free
parameters, allow us to create an efficient and accurate
numerical algorithm. 

We study in detail the phase diagram for anisotropy parameters
typical of YBCO and BSCCO. For YBCCO our results are essentially the
same as those obtained by Buzdin and Simonov, except very close to the
lower critical field. For BSCCO our results are new.

The main  results of this paper are  summarized
in the  generic phase diagram show in Fig.\ \ref{fig.phd}. We find
three  distinct phases, hereafter called phase-I, phase-II, and
phase-III. We  find that phase-I is new and  closely
related to the vortex-line  equilibrium state at the lower critical field. We also   find that phase- II is essentially identical to that found by Buzdin and Simonov
and that phase-III is   that
expected for  fields much larger than the lower critical field.
In this limit the VLL unit cell 
is related to  that appropriate for the isotropic superconductor by
scaling, and ${\bf B}\simeq {\bf H}$ ($\theta \simeq \alpha$)~\cite{cdk}.

\begin{center}
\begin{figure}
\epsfig{file=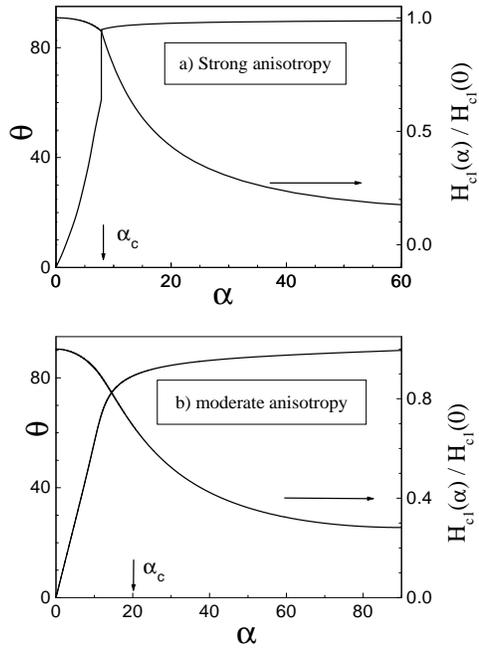,height=10cm,width=7.07cm,clip=}
\caption{ Results of Ref.[5] for
$H_{c1}(\alpha)$ and $\theta$ vs. $\alpha$ for: a) strong anisotropy,
b) moderate anisotropy} 
\label{fig.hc1}
\end{figure}
\end{center}

The lower critical field in isotropic superconductors  is independent
of the orientation $(\alpha)$, and has magnitude $H_{c1}$. 
For $H$ just above $H_{c1}$, the vortex state corresponds to vortex
lines parallel to ${\bf H}$ very far apart. In our
notation this state is described by  $\theta=\alpha$ and $L_1$, $L_2 =
\infty$. 
In a superconductor with uniaxial anisotropy, the magnitude of the lower critical field depends on $\alpha$, hereafter denoted  by $H_{c1}(\alpha)$
\cite{sbh}. The  
vortex state for $H$ just above $H_{c1}(\alpha)$ consists of vortex
lines not parallel to ${\bf H}$ ($\theta\neq\alpha$) in one of two possible
configurations, named dilute vortex line state (DVL) and dilute
vortex-line chain state (DVLC)~\cite{wdg}. The DVL is similar to
the traditional vortex state for the isotropic superconductor, in the
sense that  $L_1$, $L_2= \infty$, whereas the DVLC corresponds to
vortex-line chains along the x-direction with period $L_1$, very far
apart, that is   $L_1$ finite and $L_2 = \infty$. The latter
possibility  results from attractive interactions between vortex lines
that only exist in anisotropic superconductors
\cite{gris,vatr1,vatr2,vatr3}. To decide if the DVL or the DVLC is the 
equilibrium vortex state it is necessary to calculate
$H_{c1}(\alpha)$ for each one. The equilibrium configuration is that
with the smallest $H_{c1}(\alpha)$.
This calculation is reported  in Ref.\cite{wdg}. The results for
anisotropy parameters typical of BSCCO and YBCO, referred to as
strong an moderate anisotropies, respectively, are shown in Fig.\
\ref{fig.hc1}. 
It is found that for both anisotropies there is a value of the external 
field tilt angle, $\alpha_c$, around which the equilibrium vortex states
undergoes a significant change. For $\alpha>\alpha_c$ and  both for strong
and moderate anisotropies the equilibrium state is a DVL with vortex
lines tilted at $\theta \simeq 90^o$, that is parallel to the ab-plane.
For strong anisotropy the vortex 
state for $\alpha<\alpha_c=7.9^o$ is a DVLC with vortex lines tilted at
$\theta<60^o$. At $\alpha=\alpha_c$, there is coexistence of a DVL and
DVLC. A discontinuous change in $\theta$ takes place at $\alpha=\alpha_c$.
For a moderate anisotropy, $\theta$ is found to change smoothly with
$\alpha$ and $\alpha_c\sim 20^o$. The calculation on Ref.\cite{wdg} finds
that for moderate anisotropy the DVL and DVLC values of
$H_{c1}(\alpha)$ are so close  that their difference is
within the numerical errors of the calculation.

We find that phases I and II evolve, respectively, from the $\alpha <
\alpha_c$ and $\alpha > \alpha_c$ vortex states at $H=H_{c1}(\alpha)$,
represented by the horizontal line in Fig.\ \ref{fig.phd}.

Phase-II is essentially an extension into the  $H>H_{c1}(\alpha)$
region of the phase diagram of the vortex state found for $\alpha > \alpha_c$
at $H=H_{c1}(\alpha)$. Its predominant characteristic is that $\theta
\sim 90^o$ and, consequently, the VLL within phase-II is  essentially
identical to that for vortex lines parallel to the ab-plane. 
In this case  the unit cell is, approximately, related to that 
for the isotropic superconductor by 
scaling relations that depend only on the anisotropy. For strong and
moderate anisotropies the unit cell is a rectangle elongated in the 
y-direction, that is $L_1\gg L_2$. Phase-II
is limited from above by the line $H=H^{*}(\alpha)$, which gives the
field value around which $\theta$ changes from $90^o$ to a lower value,
accompanied by significant changes in the VLL unit cell. At
$\alpha=\alpha_c$, $H^{*}(\alpha)=H_{c1}(\alpha)$.  

Phase-I is the region of the phase diagram where the VLL properties
are strongly influenced  by the attractive interactions between pairs of
vortex lines aligned along the x-directions. 
For both strong and moderate anisotropies we find that phase-I is not
limited to $\alpha < \alpha_c$, but it also exists in a small part of
the region $\alpha > \alpha_c$ and $H>H^{*}(\alpha)$. For strong
anisotropy we find that phase-I consists of vortex-line chains along
the x-direction, with a period $L_1$ that is nearly
identical to that of an  isolated chain with the same tilt angle (which
differs from that of an isolated vortex),
separated from each other by $L_2\gg L_1$. This means that the
attractive interactions between vortex lines prevail in determining the
VLL periodicity along the x-direction.   We also find that as $H\rightarrow
H_{c1}(\alpha)$  from above, the equilibrium VLL  approaches the same
DVLC found in Ref.\cite{wdg} for $H=H_{c1}(\alpha)$. For moderate anisotropy we
obtain a new result for the vortex state at $H=H_{c1}(\alpha)$. By
extrapolating our data  as $H\rightarrow H^{*}(\alpha)$  from above, we
conclude that  the vortex state at $H=H_{c1}(\alpha)$ is a DVLC. However,
within phase-I for moderate anisotropy, the attractive interactions
between vortex lines are so weak that  vortex-line chains along the
x-direction with periods close to those of isolated chains only exist
within phase-I near  $H=H_{c1}(\alpha)$. However, distinctive
behavior, due to attractive vortex-line interactions, can still be
identified within phase-I.

This paper is organized as follows. In  Sec.\ \ref{sec.lt} we review
London theory for the 
energy of  the vortex-line system under consideration and the known
results for the phase diagram at $H=H_{c1}(\alpha)$ and for particular
cases at 
$H>H_{c1}(\alpha)$. In Sec.\ \ref{sec.numme} we discuss the
numerical method used to minimize the Gibbs free energy. In
Sec.\ \ref{sec.res}  
we report the results of this minimization, the detailed properties
of each phase, and discuss how the phases are identified. Our
conclusions are stated in Sec.\ \ref{sec.concl}. In the Appendix 
the relevant formulas for the fast convergent series used in our 
in our calculation are summarized.

\section{Method of Calculation} 
\label{sec.moc}

Here, we review London theory for this problem, its known
results for the phase diagram, and  describe some details of our
numerical method.

\subsection{London Theory}
\label{sec.lt}

To obtain the zero-temperature phase diagram for a superconductor in an
applied field  it is necessary to  determine  the 
vortex arrangement that minimizes the Gibbs free-energy density 
\begin{equation}
G=F - \frac{{\bf H}\cdot{\bf B}}{4\pi} \;
\label{eq.gfe1}
\end{equation}
with ${\bf H}$ and the volume kept constant. In Eq.\ (\ref {eq.gfe1})
$F$ is the vortex energy density and ${\bf B}$ is the magnetic
induction. We assume that the zero-temperature
equilibrium vortex configuration  consist of
straight vortex lines, parallel to each other,  arranged on a periodic
VLL, with a single vortex line 
per primitive unit cell. In this case  ${\bf B}$ is parallel to the
vortex lines direction and has magnitude $B=\Phi_0/A_c$, where
$A_c$ is the VLL
primitive unit cell area. Minimization of $G$ gives then the
vortex lines orientation and unit cell shape that corresponds to
equilibrium. Because $F$ is unchanged, if the vortex lines are  rotated
around the $c$-axis, the equilibrium ${\bf B}$ lies in the same plane as
${\bf H}$ and the $c$-axis. Thus, for ${\bf H}$ with magnitude $H$ and
tilted with respect to the  $c$-axis  by an angle $\alpha$, $G$ depends
on the vortex lines tilt angle with respect to the $c$-axis, $\theta$, and
on the parameters defining the VLL primitive unit cell. 
In the system of axis defined in Fig.\ \ref{fig.hbcu}.a, the
generic VLL primitive unit cell  lies in the $x-y$ plane,
and is defined by the primitive unit vectors ${\bf L_1}$ and ${\bf L_2}$,
as shown in  Fig.\ \ref{fig.hbcu}.b. In this case
$A_c=L_1L_2\sin{\varphi}$.

In the London limit the superconductor is
characterized by the penetration depths $\lambda_{ab}$ and
$\lambda_c$,  for currents parallel to the $ab$-plane and to the
$c$-direction, respectively. The free energy density 
of a generic arrangement of these vortex lines can be written as the
sum of pairwise interactions~\cite{rev2}  
\begin{equation}
F= \frac{\Phi^2_0}{8\pi A}\sum_{i,j} f({\bf r}_i-{\bf r}_j) \; , 
\label{eq.fre}
\end{equation}
where ${\bf r}_i$ is the i-th vortex-line  position vector in $x-y$
plane, $A$ is the sample area in this plane, and   $f({\bf r})$ is the
Fourier transform of 
\begin{equation}
f({\bf k})= e^{-2g(\bf k)}\;
\frac{1+\lambda^2_{\theta}k^2}{(1+\lambda^2_{ab}k^2)
(1+\lambda^2_{\theta}k^2_x+\lambda^2_c k^2_y)}\; ,
\label{eq.fk}
\end{equation}
where 
\begin{equation}
\lambda^2_{\theta}=\lambda^2_{ab}
\sin^2{\theta}+\lambda^2_c\cos^2{\theta} \; ,  
\label{eq.lant}
\end{equation}
and $g(\bf k)$ is the vortex core cutoff function~\cite{rev2,cdk}.
It is costumery to write $G$  as 
\begin{equation}
G(\theta,{\bf L_1}, {\bf L_2}; {\bf H})=\frac{B}{\Phi_0}[\varepsilon_{\rm
sf}(\theta)+ \varepsilon_{\rm in}(\theta,{\bf L_1},{\bf L_2})
-\frac{\Phi_0}{4\pi}H\cos{(\theta-\alpha)}] \; , 
\label{eq.gfe2}
\end{equation}
where $\varepsilon_{\rm sf}(\theta)$ is the vortex line self-energy and
$\varepsilon_{\rm in}(\theta,{\bf L_1},{\bf L_2})$ is the interaction energy 
per vortex line. In terms of  $f({\bf r})$, defined in Eq.\ (\ref
{eq.fre}), these quantities are given by
\begin{equation}
\varepsilon_{sf}(\theta)=\frac{\Phi_0}{8\pi} f({\bf r}=0) \; ,
\label{eq.esf1}
\end{equation}
and
\begin{equation}
\varepsilon_{in}=\frac{\Phi_0}{8\pi}\sum_{i\neq j} 
f({\bf R}_i-{\bf R}_j) \; ,
\label{eq.ein1}
\end{equation}
where ${\bf R}_i$ denotes the VLL positions (${\bf R}_i=n_1{\bf L}_1 +
n_2{\bf L}_2$, $n_1$, $n_2=$ integer).

The minima of $G$, Eq.\ (\ref {eq.gfe2}), are
known in the following cases. 

i) {\it Lower critical field:} $H=H_{c1}(\alpha)$. The equilibrium
vortex phase at the lower critical field is studied by Sudb\o, Brandt and Huse
\cite{sbh},   assuming that it consists of a dilute  
arrangement of straight vortex lines (DVL). Accordingly, these authors
minimize $G$, Eq.\ (\ref {eq.gfe2}), neglecting $\varepsilon_{\rm in}$,
and using for $\varepsilon_{\rm sf}$ the expression derived by
Sudb\o, and Brandt~\cite{sb}, based on an elliptic core cutoff function, 
\begin{equation}
\varepsilon_{\rm sf}(\theta)= \varepsilon_0 \frac{\lambda_{\theta}}{\lambda_c}
[\ln({\kappa}/{\gamma}) + \frac{\lambda^2_c \cos^2{\theta}}
{\lambda^2_c \cos^2{\theta}+\lambda^2_{\theta}}\ln{\frac{\gamma^2(\lambda^2_c
+\lambda^2_{\theta})}{2\lambda^2_{\theta}}}] \; ,   
\label{eq.esf2}
\end{equation}
where  $\varepsilon_0=(\Phi_0/4\pi\lambda_{ab})^2$,
$\gamma=\lambda_{ab}/\lambda_c=\xi_{c}/\xi_{ab}$,
$\kappa=\lambda_{ab}/\xi_{ab}$,   
$\xi_{ab}$ being the $ab$-plane coherence length. This line energy is
obtained from the anisotropic Ginzburg-Landau theory using the
Klemm-Clem transformation~\cite{sb,kc}. The elliptical
core cutoff has semi major axis $\xi^{-1}_{ab}$  and semi minor axis
$\xi^{-1}_{\theta}$ with $\xi^{2}_{\theta}=\xi^{2}_{ab}\cos^2{\theta} +
\xi^{2}_{c}\sin^2{\theta}$.  
Sudb\o, Brandt and Huse~\cite{sbh} obtain $H_{c1}(\alpha)$ for several
values of the anisotropy parameter $\gamma$ and of $\kappa$, and find
that for certain ranges of $\gamma$ and $\kappa$, there is a value of
$\alpha$, $\alpha=\alpha_c$, for which coexistence of  two DVL
differing from each other by the equilibrium $\theta$ takes place.

In a recent publication by some of us~\cite{wdg}, the Sudb\o, Brandt and
Huse~\cite{sbh} calculation was generalized to account for two possible
vortex phases at the lower critical field: a dilute vortex-line state
(DVL) considered by 
them and  a  dilute  vortex-line chains arrangement (DVLC). The DVLC
consists of vortex lines tilted with respect to the $c$-axis, parallel to
one another, and aligned  along the x-direction, forming a periodic
chain. This 
is possible because the interaction between a pair of such vortices is
attractive at large distances~\cite{gris,vatr1,vatr2,vatr3}. 
In Ref:\cite{wdg}, $G$, Eq.\ (\ref {eq.gfe2}), is minimized in the limit
of vanishing vortex density, for each one of these possibilities. For
the DVL, $\varepsilon_{\rm int}$ is neglected in Eq.\
(\ref {eq.gfe2}), whereas for the  DVLC $\varepsilon_{\rm int}$ in Eq.\
(\ref {eq.gfe2})  is identified with the interaction energy per vortex
line of an isolated chain. The lower critical field as a function of
the  tilt angle $\alpha$,  $H_{c1}(\alpha)$ is then obtained for each
possibility, and the equilibrium vortex phase at $H=H_{c1}(\alpha)$  is
identified as that with the smallest $H_{c1}(\alpha)$.

In Ref:\cite{wdg}, two sets of anisotropy parameters are considered:

\noindent {\it Strong anisotropy}: $\kappa=10$, $\gamma=1/\surd{200}$

\noindent {\it Moderate anisotropy}: $\kappa =50$, $\gamma=1/5$

\noindent These values for $\gamma$  are typical of YBCO (moderate
anisotropy) and BSCCO 
(strong anisotropy).  The parameter $\kappa$ for
moderate anisotropy is typical of YBCO but for strong anisotropy is
about five times smaller than those typical of BSCCO~\cite{rev1}.
However, this difference does not alter significantly the phase diagram
because only the self-energy depends on $\kappa$. 

The main results of Ref.\cite{wdg}, summarized in Fig.\ \ref{fig.hc1} are as
follows. For both moderate 
and strong anisotropy and for  $\alpha>\alpha_c$ the vortex lines are
nearly parallel to the a-b plane ($\theta\sim 90^o$). For strong
anisotropy $\alpha_c=7.9^o$ and for moderate anisotropy $\alpha\sim 20^o$.
For strong anisotropy and $\alpha<\alpha_c$ the vortex phase at $H= 
H_{c1}(\alpha)$ is a DVLC. For $\alpha>\alpha_c$, on the other hand, 
no significant difference is found between the DVLC and DVL phases,
because the vortex lines are tilted  at $\theta \sim 90^o$, and at such
tilt angles the equilibrium chain period is very large. Phase coexistence
is also found for strong anisotropy at $\alpha=\alpha_c=7.9^o$, but the
coexisting phases are a  DVLC and  DVL, instead of the two DVL found in 
Ref.\cite{sbh}. A discontinuous jump in $\theta$
takes place as $\alpha$ crosses $\alpha_c=7.9^o$.
For moderate anisotropy the DVL and DVLC phases give nearly identical
values for $H_{c1}(\alpha)$, so that no  conclusion on which is the
equilibrium state is reached.

ii) $\alpha=0^o$ $and$ $\alpha=90^o$. In both cases 
 $\theta=\alpha$ and the lower critical field  is related to
the self energy by  the usual formula
$H_{c1}(\alpha=0^o,90^o)=4\pi\varepsilon_{\rm
sf}(\theta=0^o,90^o)/\Phi_0$. For $\alpha=\theta=0^o$ the equilibrium VLL is
the  familiar isotropic superconductor triangular lattice,  with
$L_1=L_2=L_{\Delta}\equiv (\Phi_0/B\sin{60^o})^{1/2}$, $\varphi=60^o$ 
and $\psi$ undetermined. For $\alpha=\theta=90^o$  the x-direction
coincides with the negative $c$-axis. The equilibrium  VLL is related to
that for $\alpha=\theta=0^o$ (assuming ${\bf L}_1$ along the x-direction or
$\psi=0^o$)  by the following scaling relations~\cite{rev1} 
\begin{eqnarray}
{\bf L}_1 & = & L_{0}\gamma^{1/2}
{\bf \hat{x}}  \nonumber\\
{\bf L}_2 & = & L_{0}[\gamma^{1/2}
\cos{\varphi} {\bf \hat{x}}+ \gamma^{-1/2}
\sin{\varphi} {\bf \hat{y}}] \; ,
\label{eq.ph90}
\end{eqnarray}
where $L_0=(\Phi_0/B\sin{\varphi})^{1/2}$, with $\varphi=60^o$ for the
triangular lattice. These results also apply if the $\alpha=0^o$ VLL is
a square lattice, which will be considered in Sec.\ \ref{sec.res}, in which
case $\varphi=90^o$ in Eq.\ (\ref {eq.ph90}).  
The unit cell defined by Eqs.\ (\ref {eq.ph90}) results from
compressing the  $\alpha=\theta=0^o$ VLL unit cell (with ${\bf L}_1$
along the negative $c$-direction) by $\gamma^{1/2}$ along the $c$-direction and
stretching it  by $\gamma^{-1/2}$ in the ab plane. This transformation
conserves the unit cell area $A_c=L^2_0\sin{\varphi}$. The $B$ vs. $H$
curves for $\theta=\alpha=0^o$ and $\theta=\alpha=90^o$  are also
related by scaling as follows
\cite{rev1,rev2} 
\begin{equation}
B_{90^o}(\frac{H-H_{c1}(90^o)}{H_{c1}(0^o)})=\gamma
B_{0^o}(\frac{H-H_{c1}(0^o)}{\gamma H_{c1}(0^o)}) \; .
\label{eq.bscl}
\end{equation}  
 
iii) High fields ($H\gg H_{c1}(\alpha)$). To a good degree of
approximation the equilibrium VLL is related to the $\alpha=\theta=0^o$
VLL by the   scaling relations 
Ref.\cite{cdk}  
\begin{eqnarray}
{\bf L}_1 & = & L_{0}(\frac{\lambda_{\theta}}{\lambda_c})^{1/2}
\; {\bf \hat{x}}  \nonumber\\
{\bf L}_2 & = & L_{0}[(\frac{\lambda_{\theta}}{\lambda_c})^{1/2}
\cos{\varphi}\; {\bf \hat{x}}+(\frac{\lambda_c}{\lambda_{\theta}})^{1/2}
\sin{\varphi} \; {\bf \hat{y}}] \; .
\label{eq.hih}
\end{eqnarray}
In this case  ${\bf B}\simeq {\bf H}$, with
$B=\Phi_0/L^2_{\Delta}\sin{\varphi}$, with $\varphi=60^o$ for the
triangular VLL and $\varphi=90^o$ for the square VLL.

Apart from these cases, little else is known about the vortex phase
diagram. Results for $H=H_{c1}(\alpha)$, such as the occurrence of a
DVLC as the equilibrium state, phase coexistence, and 
large differences between  $\theta$ and $\alpha$,  suggest a  
low field ($H\gtrsim H_{c1}(\alpha)$) phase diagram that is non-trivial
and that  differs considerably from the high-field one. It is of
interest then to obtain a full picture of the phase diagram. This is
our main motivation here.

According to the discussion above, for a given ${\bf H}$, $G$ depends
on five independent variables, $\theta,{\bf L_1},{\bf L_2}$, or,
equivalently, on $\theta, L_1, L_2,\varphi,\psi$ (Fig.\
\ref{fig.hbcu}.b). It can be shown that the  dependence of $G$ on the
angle of rotation of the VLL with respect to ${\bf B}$, $\psi$, is such
that $G$ has a minimum for $\psi=0$. We assume that this is the
absolute minimum and, from here on, take $\psi=0$ in our calculations.
In order to simplify the calculations, we fix the value of the angle
$\varphi$ at  $\varphi=90^o$. Thus, in the results reported here, ${\bf
L_1}$ is along the x-axis and coplanar with ${\bf B}$, ${\bf H}$ and
the $c$-axis,  ${\bf L_2}$ is along the y-axis, and the VLL unit cell is
rectangular. We find that setting $\varphi=90^o$ does not change the general
picture for the phase diagram. With these restrictions $G$ depends
only on three variables: $\theta, L_1, L_2$. In the remainder of the
paper we use instead $\theta, L_1, B=\Phi_0/(L_1L_2)$ as independent variables.

One novel aspect of our calculation is to account for the effects 
of interactions between vortex lines in the zero-temperature phase
diagram.  We find that the simple model described above has a
non-trivial low-field phase diagram because attractive interactions
between vortex lines give an important contribution 
to $\varepsilon_{\rm int}$ at low fields. The reason is that, as
discussed above, the equilibrium VLL unit cell has ${\bf L_1}$ along the
x-direction ($\psi=0$). This 
means that the equilibrium phase consists of periodic vortex-line chains   
aligned along x, with period $L_1$, separated from each other by $L_2$.
According to~\cite{gris,vatr1,vatr2,vatr3}, the interaction between a pair of
vortex lines in the same chain is attractive if their separation $x$ is
greater than $x_m(\theta)$, and repulsive if  $x<x_m(\theta)$. Thus, as
long as $L_1$ is not too small compared to  $x_m(\theta)$, the
vortex-line chains contribution to $\varepsilon_{\rm int}$ is
predominantly attractive and essentially negative.
Interactions between interchain vortex lines are always repulsive, thus
giving a positive contribution to $\varepsilon_{\rm int}$. 

In the DVLC state at $H=H_{c1}(\alpha)$, these
chains are infinitely far apart ($L_2=\infty$) and $L_1$ minimizes
$\varepsilon_{\rm int}$ for the equilibrium $\theta$. In this paper, we
obtain the vortex phases at $H=H_{c1}(\alpha)$ by extrapolating our
results for $H>H_{c1}(\alpha)$. For strong anisotropy, our results are in
agreement with those of Ref.\cite{wdg}. For moderate anisotropy, we
find that the vortex phase at $H=H_{c1}(\alpha)$ is also a DVLC for
$\alpha<\alpha_c$, a result not obtained in Ref.\cite{wdg}.

Our results for the zero-temperature phase diagram  can be  summarized
in the  generic phase diagram show in Fig.\ \ref{fig.phd}, consisting
of three distinct phases, whose main characteristics are discussed in
Sec.\ \ref{sec.int}.

\subsection{Numerics}
\label{sec.numme}

The main difficulty to  carry out the minimization of $G$, Eq.\ (\ref
{eq.gfe2}), is  to evaluate the lattice sum in the expression for 
$\varepsilon_{\rm in}$, Eq.\ (\ref {eq.ein1}). This can only be done
numerically. However, the direct evaluation of this sum 
 is impractical for our purposes,
because it converges very slowly. To circumvent this difficulty we use
the results obtained by Doria~\cite{mmd}, who  showed
that the lattice sum in  Eq.\ (\ref {eq.ein1}) can be transformed into
a series that converges much faster than the direct sum. 

Our numerical method uses a simulated annealing algorithm to locate
the minima of $G$. For a given ${\bf H}$, specified by $H$ and $\alpha$,
the minimization of $G$ is carried out in the space spanned by $\theta,
L_1$, and $B$. Our algorithm starts from a convenient choice for these
variables and attempts to change them to new values that decrease $G$.
Once a set of such values is found, another one is searched by
repeating the procedure, and so on. This converges to a set
corresponding to a minimum of $G$ after several steps.  
For each new set of $\theta, L_1,B$  we  calculate
$\varepsilon_{\rm in}$,  using Doria's  fast convergent series. This
allows us to  run the simulated annealing for a very
large number of steps, and to obtain the minima of $G$ with high
accuracy, except when $H$ is very close to  $H_{c1}(\alpha)$. In this
case our method converges  slowly, because the VLL unit cell becomes
very large and $\varepsilon_{\rm in}$  very small. A similar numerical
method is described in Ref.\cite{wg}

\section{Results}
\label{sec.res}

Here we report in detail the results obtained by the numerical method
described in Sec.\ \ref{sec.numme} for strong and moderate anisotropies.
To explore the ($H,\alpha$) phase diagram   we fix  $\alpha$
and obtain the minima of $G$ for several values of
$H>H_{c1}(\alpha)$. By repeating this procedure for several $\alpha$ we
obtain the phase diagram.  We check  that these minima  are  absolute ones
by running the above described simulated annealing algorithm starting from distinct initial states, and selecting  the minima with the smallest $G$.

\subsection{Strong anisotropy}

\begin{center}
\begin{figure}
\epsfig{file=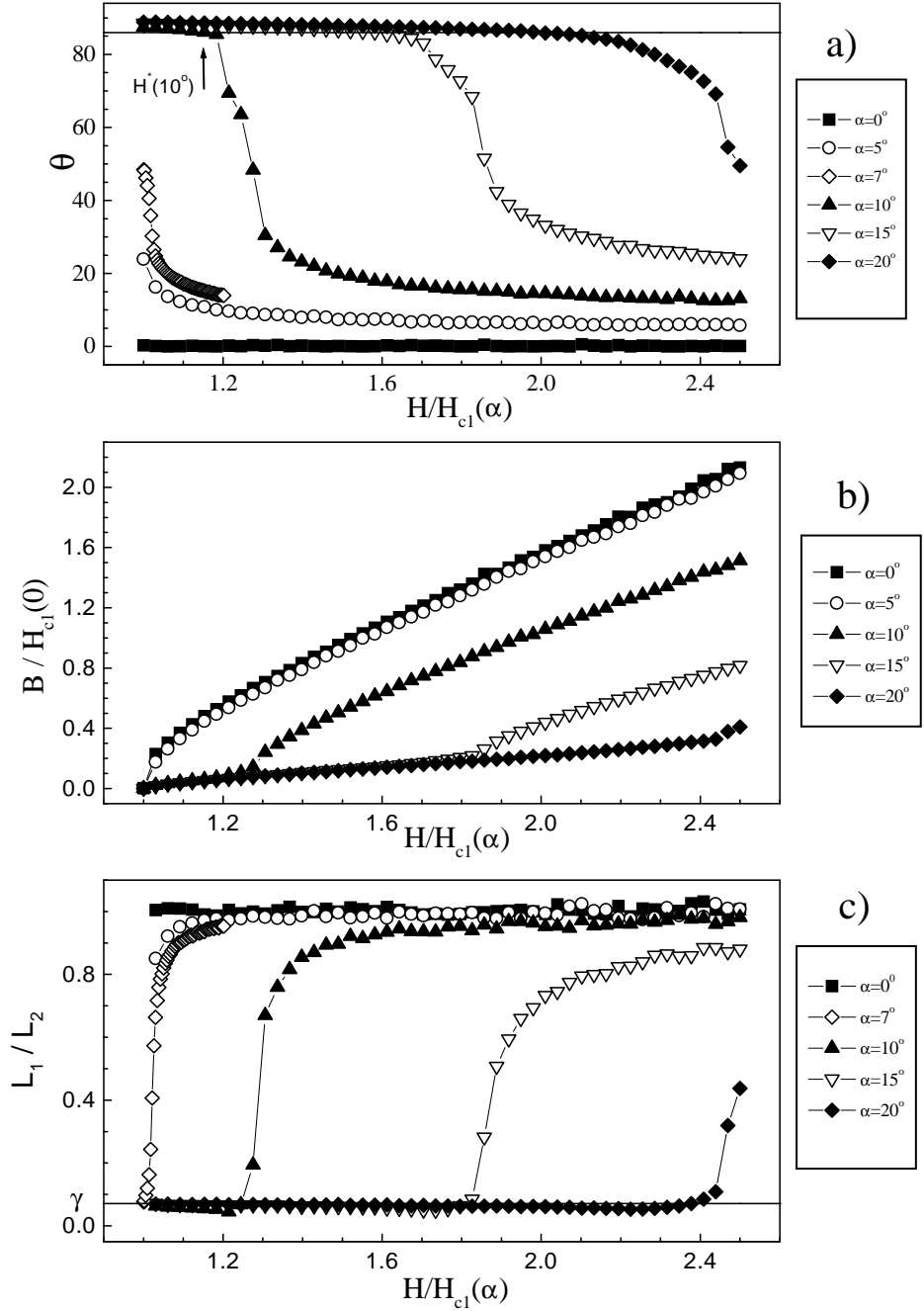,height=20cm,width=14cm,clip=}
\caption{ Numerical results for strong anisotropy. a) Vortex lines tilt
angle. The full line $\theta=86^o$ corresponds to
$\tan{\theta}=1/\gamma$. The arrow indicates $H^{*}(\alpha=10^o$). b)
Magnetic induction. c) Ratio between VLL unit cell sides. The full line
$\L_1/L_2=\gamma$ corresponds to the value predicted by scaling for
$\alpha=90^o$ }
\label{fig.stga}
\end{figure}
\end{center}

\begin{center}
\begin{figure}
\epsfig{file=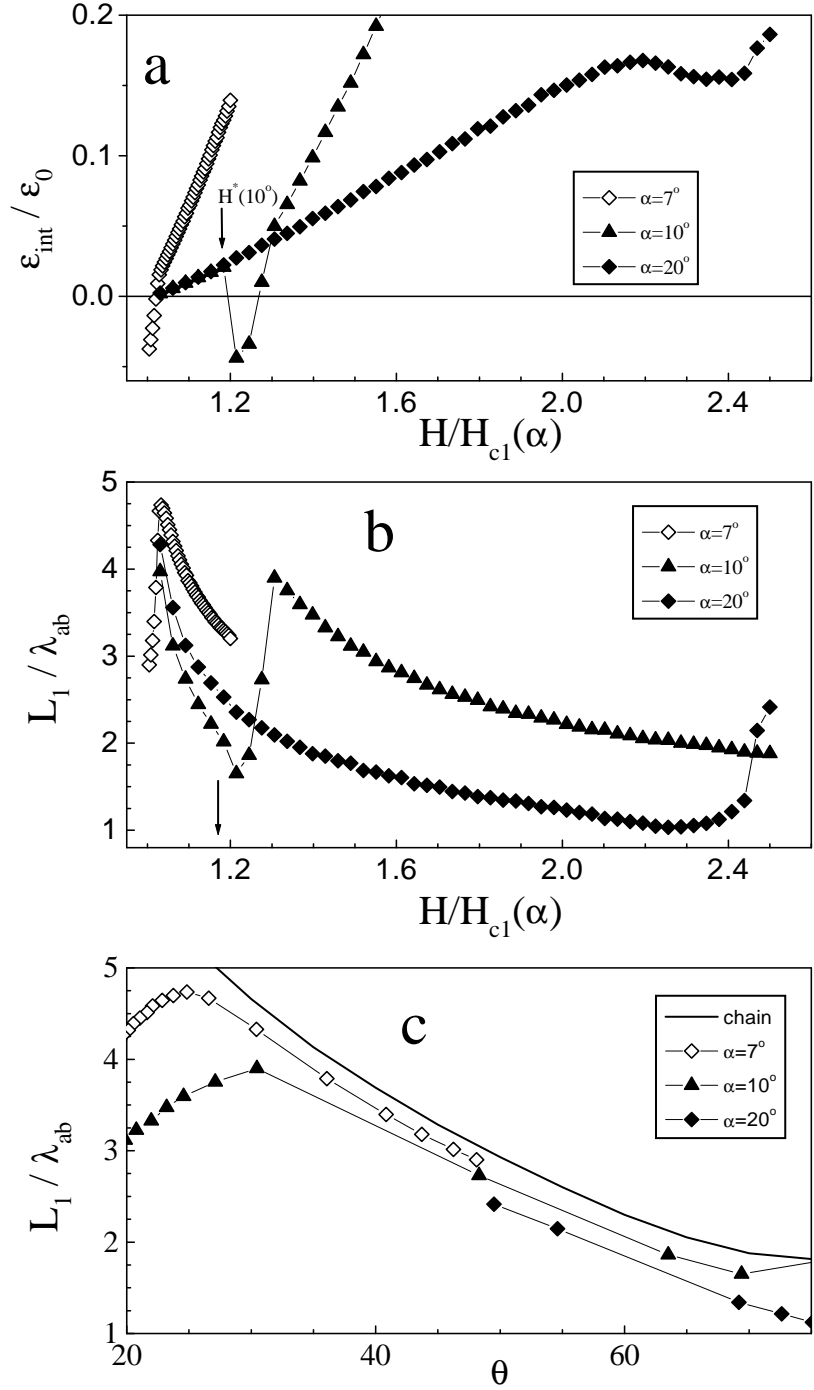,height=20cm,width=14cm,clip=}
\caption{ Numerical results for strong anisotropy. a) Interaction
energy per vortex line. Arrow indicate the value of $H^*$ for
$\alpha=10^o$. b) VLL period along x. c) VLL  period along x vs.
tilt angle compared with the period of an
isolated vortex-line chain tilted at the same angle (full curve) }
\label{fig.stglec}
\end{figure}
\end{center}

The main results are summarized  in Figs.\ \ref{fig.stga} and
\ref{fig.stglec}.  

First, we argue  that our results   are consistent with the prediction,
discussed in Sec.\ \ref{sec.int}, that at $H=H_{c1}(\alpha)$ the vortex
configuration for $\alpha<\alpha_c=7.9^o$ is a DVLC.  

The existence of a DVLC   at $H=H_{c1}(\alpha)$ for $\alpha<\alpha_c$
means that as $H/H_{c1}(\alpha)\rightarrow 1$ the primitive unit
cell sides behave as $L_1\rightarrow L_{ch}$(= chain period),
$L_2\rightarrow\infty$, and that  $\varepsilon_{in}$, Eq.\ (\ref
{eq.gfe2}), approaches a negative value, equal to the  interaction
energy per vortex line in an isolated chain.  
Our results  for $\alpha =7^o$ show exactly this behavior. The
$L_1/L_2$ curve in Fig.\ \ref{fig.stga}.c  
decreases  continuously  as $H/H_{c1}(\alpha)\rightarrow 1$, indicating that 
$L_1/L_2$ extrapolates to $L_1/L_2=0$ at $H=H_{c1}(\alpha)$. On the
other hand, the $L_1$ curve in (Fig.\ \ref{fig.stglec}.b) approaches a
finite value . We find that the extrapolated values of $L_1$, 
$\theta$ and $\varepsilon_{in}$  at $H=H_{c1}(\alpha)$   are
$L_1/\lambda_{ab}=2.8$, $\theta=50^o$  and
$\varepsilon_{in}/\varepsilon_0=-0.04$. These values 
 agree, within numerical errors, with the chain period,  tilt angle and
interaction energy per vortex line of the isolated vortex-line chain 
obtained in Ref.\cite{wdg}. 

Next we consider the  behavior for $\alpha<\alpha_c$ and
$H>H_{c1}(\alpha)$. 

\begin{center}
\begin{figure}
\epsfig{file=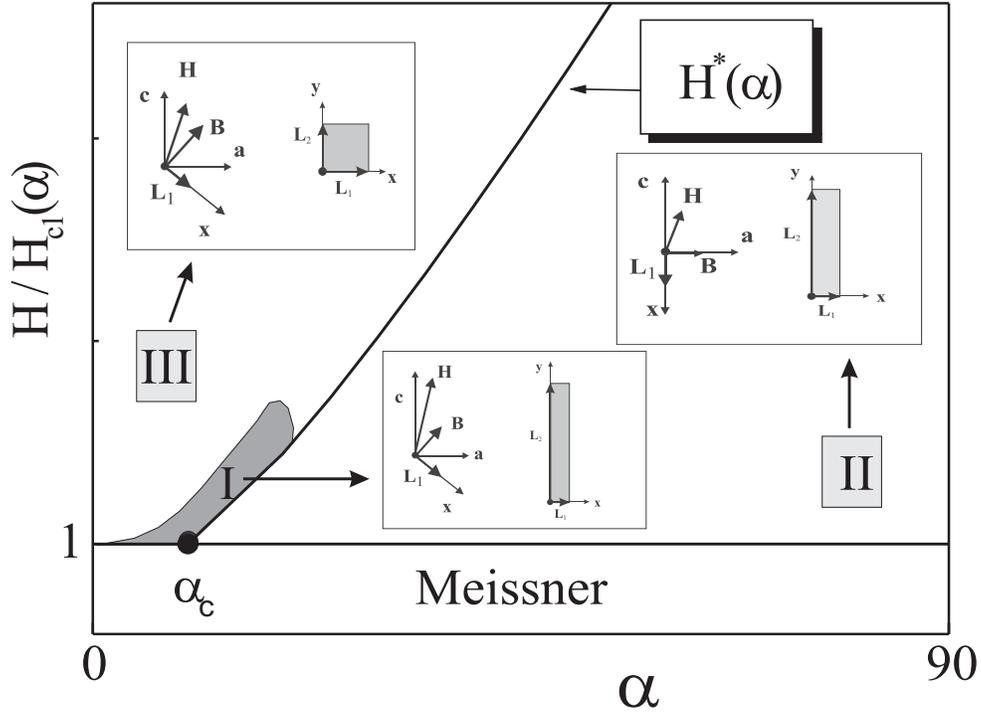,height=10 cm,width=14.15 cm,clip= }
\caption{ Generic phase diagram in the ($H,\alpha$) plane for both
strong and moderate anisotropy.}
\label{fig.phd}
\end{figure}
\end{center}

For  $\alpha =7^o$  our results for the $L_1$ vs. $H$  curve (Fig.\
\ref{fig.stglec}.b) show that there are two distinct behaviors. 
One for $1<H/H_{c1}(7^o)<1.03$, where $L_1$ increases with $H$, and
another for $H/H_{c1}(7^o)>1.03$ where $L_1$ decreases with $H$.
The $\varepsilon_{in}$ vs. $H$ curve changes  slope at
$H/H_{c1}(7^o)=1.03$, and shows that $\varepsilon_{in}$ increases
continuously with $H$, being  negative for $H/H_{c1}(7^o)<1.02$.
A decrease of $L_1$  with $H$ is what is expected if interactions
between vortex lines are purely repulsive. This behavior is identified with
phase-III (Fig.\ \ref{fig.phd}) and corresponds to the region where
$L_1/L_2>0.6$ , $\theta<23^o$ and $\varepsilon_{in}>0$.  The increase
of  $L_1$ with $H$, on the other hand, arises from  attractive
interactions between vortex lines. This can be seen by plotting $L_1$
vs. $\theta$ for  $\alpha =7^o$, as shown in Fig.\ \ref{fig.stglec}.c.
This figure shows that in this range of $H$ values, $L_1$ is  nearly
equal to the period of an isolated vortex-line chain with the same
$\theta$ and has essentially the same $H$ dependence. This is only
possible if the attractive intrachain vortex-line interactions are
strong compared to interchain repulsions. Further evidence for this is
found in the negative values for $\varepsilon_{in}$. We identify the
region $1<H/H_{c1}(7^o)<1.03$ as phase-I and that for
$H/H_{c1}(7^o)>1.03$ as phase-III.
For $\alpha =7^o$ we also find that $L_2>L_1$, and that  $L_2$ decreases
with $H$ in such a way that $L_1/L_2$ increases with $H$ as shown in 
Fig.\ \ref{fig.stga}.c.

A similar behavior is expected for other values of $\alpha<\alpha_c$.
For $\alpha= 5^o$ we find that the $L_1/L_2$ curve shows a downturn as
$H/H_{c1}(\alpha= 5^o)$ decreases (Fig.\ \ref{fig.stga}.c), similar to
that for  $\alpha= 7^o$, suggesting that the limit value as
$H/H_{c1}(\alpha)\rightarrow 1$ is 
small. We also find that the limit value of $\theta$ as
$H/H_{c1}(\alpha)\rightarrow 1$ (Fig.\ \ref{fig.stga}.a) agrees well
with that obtained in 
Ref.\cite{wdg}. Although, for $\alpha\leq 5^o$, we are unable to observe
a region where $L_1$ increases with $H$ and $\varepsilon_{in}<0$, we
believe that the phase diagram for  $\alpha=5^o$ and $\alpha= 7^o$ are
similar. One difficulty with values of $\alpha$ close to $0^o$ is that
the region where phase-I exists is so close to $H/H_{c1}(\alpha)=1$
that the   accuracy of our numerical method is insufficient
to observe it.

Next  we discuss the behavior for  $\alpha>\alpha_c$ and $H>H_{c1}(\alpha)$. 

Our results show that $\theta$ remains close to $90^o$ in the  range 
$H_{c1}(\alpha)<H<H^*(\alpha)$ (Fig.\ \ref{fig.stga}.a). We find  that for 
the results in this range  the  scaling
relations  discussed in Sec.\ \ref{sec.int}. For strong anisotropy
$\gamma=1/\sqrt{200}$, so that $\tan{\theta}>1/\gamma$  for
$\theta>86^o$. Accordingly,  the field
$H^*(\alpha)$ is obtained from the $\theta$ vs. $H$ curve as 
the largest $H$  for which  $\theta>86^o$, as indicated in Fig.\
\ref{fig.stga}.a.    
For $\alpha\geq 10^o$ and $H_{c1}(\alpha)<H<H^*(\alpha)$ the 
$L_1/L_2$ and $B$ vs. $H$ curves in Fig.\ \ref{fig.stga} essentially coincide.
The common curves are $L_1/L_2=\gamma=0.07$  
and  the $B$ vs. $H$ curve for $\alpha=\theta=90^o$,  obtained from that for
$\alpha=\theta=0^o$ by the  scaling relations Eq.\ (\ref {eq.bscl}).
The behavior of $L_1$, $L_2$ and $\varepsilon_{in}$ in this region is
that expected for repulsive interactions: as $H$ increases, $L_1$ and
$L_2$ decrease, and $\varepsilon_{in}$ increase (Fig.\
\ref{fig.stga}). From these results we identify the region
$H_{c1}(\alpha)<H<H^*(\alpha)$ as phase-II. 

For $H>H^{*}(\alpha)$ and for $\alpha=10^o$,
$15^o$  and $20^o$ the tilt angle $\theta$ decreases smoothly with
increasing $H$ (Fig.\ \ref{fig.stga}.a). The behaviors of $L_1$ and
$\varepsilon_{in}$  with $H$  are not smooth. As shown in Fig.\
\ref{fig.stglec} there are regions  where  $L_1$ increases with $H$ and  the
$\varepsilon_{in}$ curve shows a dip, reaching negative values for
$\alpha=10^o$. This behavior of $L_1$ is  similar to that for
$\alpha=7^o<\alpha_c$ in phase-I region. 
We also find that for $\alpha \leq 20^o$ the VLL period along x,
$L_1$, is close that for an isolated 
vortex-line chain with the same $\theta$, and has a similar $H$
dependence (Fig.\ \ref{fig.stglec}.c). In view of these similarities we
identify the region where $L_1$ increase with $H$ as phase-I. Thus the
portion of the phase diagram occupied by phase-I extends into the
region  $\alpha>\alpha_c$ for $H>H^{*}(\alpha)$, as shown in Fig.\
\ref{fig.phd}.

Our results indicate that for strong anisotropy within  
phase-I the vortex-line chains along x have essentially the same
period as an isolated chain tilted at the equilibrium $\theta$.

We find that as $\alpha$ increases above $\alpha=20^o$ 
the region where $L_1$ increases with $H$ becomes
smaller and the dip in $\varepsilon_{in}$ decreases. Eventually these
features disappear altogether at some $\alpha$, so that the phase-I
region in the phase diagram is bounded as shown in Fig.\ \ref{fig.phd}.

Our results indicate that the transition from phase-II to phase-I is a
smooth crossover, with an intermediate region starting just above
$H^{*}(\alpha)$  and ending when $\theta\sim70^o$. In this region   the
$L_1/L_2$ and $B$ curves for $\alpha=10^o$, $15^o$  and $20^o$ 
show only small deviations from  the scaling behavior of phase-II, 
with $L_1/L_2$ smaller than  the scaling value and  the $B$
curve slightly above the scaling  one (Fig.\ \ref{fig.stga}).
Behavior characteristic of phase-I, as described above, shows up only
for  $\theta\lesssim 70^o$. The reason is  is that for 
$\theta\gtrsim70^o$   interactions between intrachain vortex lines
are predominantly repulsive because $L_1\ll x_m(\theta)$.   

For  $\alpha=10^o$, $15^o$  and $20^o$ phase-I behavior ends at the $H$
value where $L_1$ stops increasing with $H$. For larger $H$, we find
that $L_1$ 
decreases smoothly with $H$, with  $L_1/L_2\sim1$, and
that $\varepsilon_{in}$ increases with $H$ (Fig.\ \ref{fig.stga}). We
identify these regions as phase-III (Fig.\ \ref{fig.phd}).

\subsection{Moderate anisotropy} 
\begin{center}
\begin{figure}
\epsfig{file=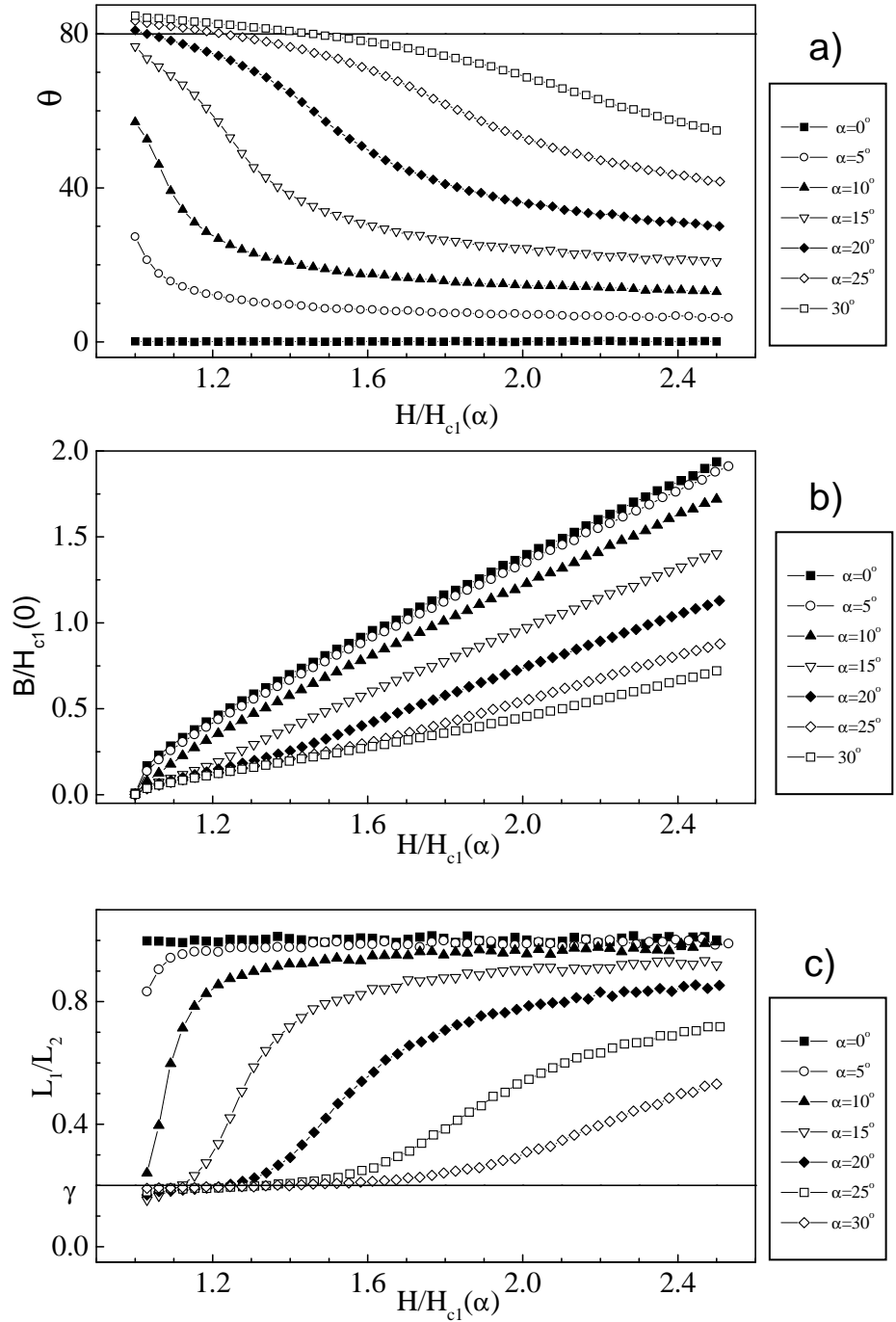,height=20cm,width=14cm,clip=}
\caption{ Numerical results for moderate anisotropy. a) Vortex lines tilt
angle. The full line $\theta=80^o$ corresponds to $\tan{\theta}=1/\gamma$. b)
Magnetic induction. c) Ratio between VLL unit cell sides. The full line
$\L_1/L_2=\gamma$ corresponds to the value predicted by scaling for
$\alpha=90^o$}
\label{fig.moda}
\end{figure}
\end{center}

\begin{center}
\begin{figure}
\epsfig{file=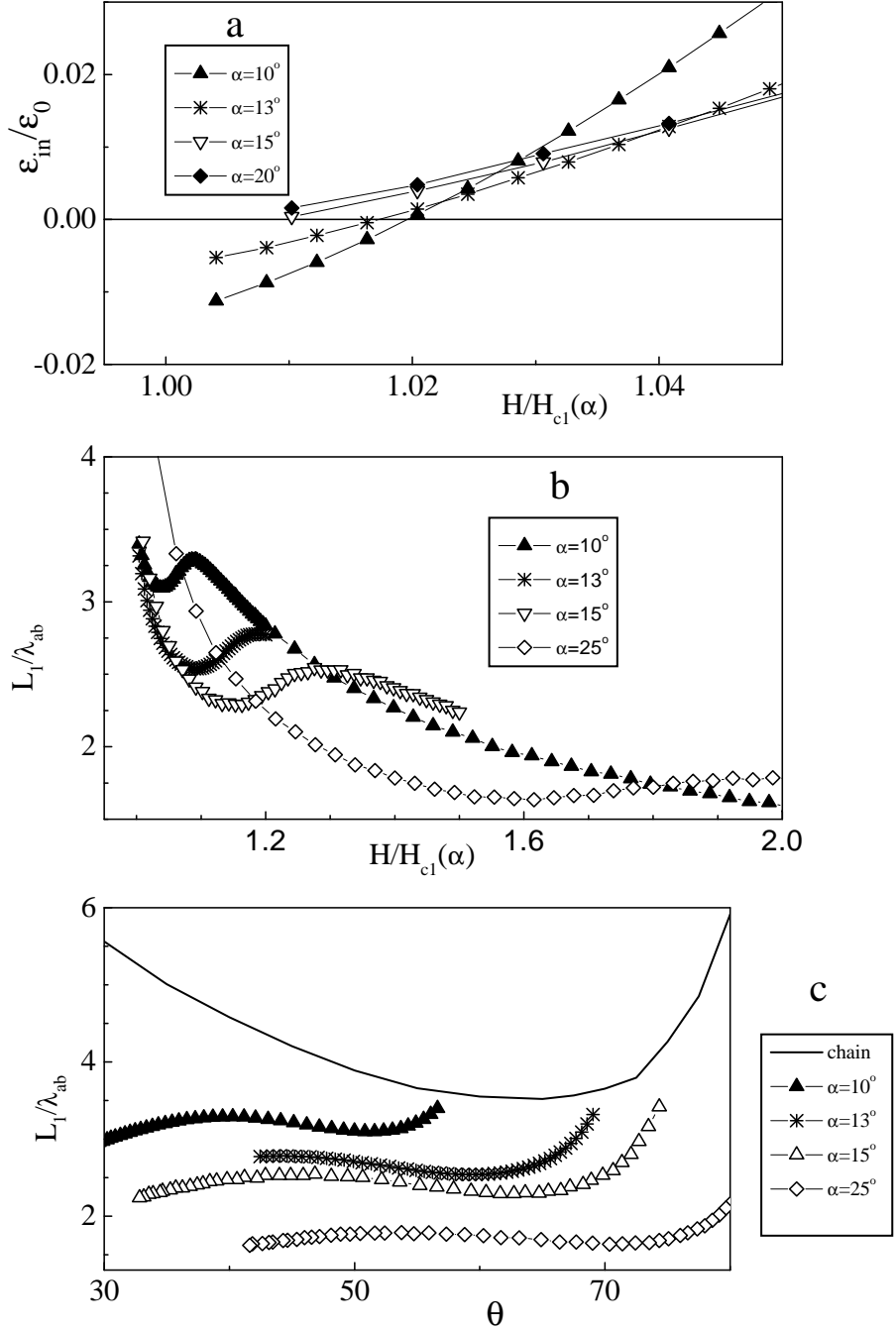,height=20cm,width=14cm,clip=}
\caption{ Numerical results for moderate anisotropy. a) Interaction
energy per vortex line.  b) VLL period along x. c) VLL
period along x vs. tilt angle compared with the period of an
isolated vortex-line chain tilted at the same angle (full curve) }
\label{fig.modlec}
\end{figure}
\end{center}

The results are similar to those
for strong anisotropy and are summarized in Figs.\ \ref{fig.moda} and
\ref{fig.modlec}.  

First we argue that as $H/H_{c1}(\alpha)\rightarrow1$ the equilibrium
state approaches a DVLC for $\alpha\lesssim20^o$.
As  seen in Fig.\ \ref{fig.moda}.c, for $\alpha=10^o$, $15^o$ and
$20^o$, $L_1/L_2$ approaches a small value as $H/H_{c1}(\alpha)\rightarrow1$,
which  we interpret as indicating that $L_2\rightarrow \infty$. The
$L_1$ vs $H$ and $L_1$ vs. $\theta$ curves, shown in Figs.\
\ref{fig.modlec}.b  and \ref{fig.modlec}.c, suggest that, as
$H/H_{c1}(\alpha)\rightarrow1$,  
$L_1$ approaches a finite value that is close to the period of 
an isolated vortex-line chain with the same $\theta$~\cite{wdg}. 
The   $\varepsilon_{in}$ vs. $H$ curve shown in Fig.\
\ref{fig.modlec}.a becomes negative as $H/H_{c1}(\alpha)\rightarrow1$
for $\alpha=10^o$ and $13^o$ and extrapolates to a negative value at
$H=H_{c1}(\alpha)$ for $\alpha=15^o$ and $20^o$. This behavior is only
possible for a DVLC state at $H=H_{c1}(\alpha)$.

Next we consider the  behavior for $\alpha<\alpha_c$ and
$H>H_{c1}(\alpha)$.

For $\alpha=10^o$, $13^o$ and
$15^o$ the $L_1$ vs. $H$ curves in Fig.\ \ref{fig.modlec}.b show
non-monotonic behavior. As $H$ increases $L_1$ first
decreases, then increases and, finally, decreases monotonically.
This behavior suggests that attractive interactions between intrachain
vortex lines are competing with repulsive interchain ones. This
 is further supported by the negative $\varepsilon_{in}$,
for the same $\alpha$ values, shown in Fig.\ \ref{fig.modlec}.b. We
interpret  as phase-I the $H$ range 
starting from  $H=H_{c1}(\alpha)$ and ending at the $H$ value in Fig.\
\ref{fig.modlec}.b where $L_1$ stops increasing. However,
the behavior in phase-I for moderate anisotropy differs from that  for
strong anisotropy. 
The $L_1$ vs. $\theta$ curve in Fig.\ \ref{fig.modlec}.c shows that
$L_1$ differs considerably from the equilibrium period of an isolated
chain with the same $\theta$ and has a different $H$ dependence.     
 
We find that for  $\alpha>\alpha_c $ our results agree with phase-II
behavior for $H_{c1}(\alpha)<H<H^*(\alpha)$.
For moderate anisotropy $\gamma=1/5$, and $\tan{\theta}>1/\gamma$  for
$\theta>80^o$. The field $H^*(\alpha)$ is obtained from the $\theta$
vs. $H$ curve as  the largest $H$  for which  $\theta>80^o$, as
indicated in Fig.\ \ref{fig.moda}.a.
We find that for $H_{c1}(\alpha)<H<H^*(\alpha)$ the $L_1/L_2$ and $B$
vs. $H$ curves shown in Fig.\ \ref{fig.moda} agree well with the
scaling predictions Sec.\ \ref{sec.int}. 

Thus we conclude that for moderate anisotropy the phase diagram is also
like that shown in Fig.\ \ref{fig.phd}.

\section{Conclusions}
\label{sec.concl}

In conclusion, we obtain the zero-temperature phase diagram for
superconductors with anisotropy parameters typical of BSCCO
and YBCO. Three distict phases are found and characterized in detail. Phase-I is where we find new results. The most significant ones are: i)in the case of strong anisotropy we determine how  the vortex-line-chain state that occurs at $H=H_{c1}(\alpha)$ ($\alpha \leq \alpha_c$)  is modified by interchain 
interactions. ii) In the case of moderate anisotropy we show that the vortex state  at $H=H_{c1}(\alpha)$ is a vortex-line chain state.
 
In phase-II, we   recover the results  of Buzdin and Simonov that reveal  a 
large region of the phase diagram over which the vortex lines remain
nearly parallel to the ab-plane. Moreover we find   
the curve, $H=H^{*}(\alpha)$, around which the vortex lines tilt away
from the ab-plane. We also find that this line starts at $\alpha \leq \alpha_c$ and  $H=H_{c1}(\alpha_c)$ and that, as this line is being crossed, the  VLL undergoes a structural change that becomes more and more  abrupt as $\alpha \rightarrow \alpha_c$.

Our results are obtained with the simplifying assumption that
$\varphi=90^o$. We have also carried out the minimization of $G$, Eq.\
(\ref {eq.gfe2}), with $\varphi$ as  independent variable. The results
are identical to those reported above, except for the VLL unit cell
shape which, in general, is not rectangular. 

Our present results seem to indicate that the equilibrium VLL changes
smoothly with $H$ and $\alpha$. In this case the phase boundaries in
Fig.\ \ref{fig.phd} are crossover lines. However, our data for strong
anisotropy shows  abrupt changes in the slope  of the curves for
$\theta$, $B$ and $L_1$ vs. $H$ for $H$ close to $H_{c1}(\alpha)$.
These could result from a rapid 
crossover or from  true discontinuities. It is found in Ref.\cite{wdg} 
that, for strong anisotropy,  if $\alpha$ is changed and $H$ is kept
along the $H/H_{c1}(\alpha)=1$ curve,  there is a
discontinuous change in $\theta$ at $\alpha=\alpha_c=7.9^o$, associated
with  coexistence of a DVLC 
state with a DVL one. This coexistence might remain for
$H/H_{c1}(\alpha)>1$ in the vicinity of $\alpha_c$. In the
present study  we do not attempt to verify  if there are true
discontinuities. However, our numerical method is capable of doing so.
Work on this topic is under way and will be reported elsewhere.

\acknowledgments
This work was supported in part by MCT/CNPq, FAPERJ, CAPES and DAAD. 
The authors thank Dr. Ernst Helmut Bandt for helpful discussions.

\appendix
\section*{}
\label{sec.app}

In this appendix the formulas that we use to calculate
$\varepsilon_{in}(\theta,L_1,L_2,\varphi)$ numerically are summarized.

Doria showed that the lattice sum in Eq.\ (\ref {eq.ein1}) (with ${\bf
L_1}$ along x) can be transformed in a  fast convergent series~\cite{mmd}.
The results are:
\begin{eqnarray}
\varepsilon_{in} &=& \big(\frac{\Phi_0}{4\pi\lambda_{ab}}\big)^2 \lbrace
\frac{\lambda_{\theta}}{\lambda_c}V_0(0) - 
(1-\gamma^2) \cos^2\theta\int^{1}_{0}
\frac{du}{\sqrt{c_1(u)c_2(u)}} 
\lbrack\, -\mu_0^2(u) \frac{\partial V_0(u)}{\partial \mu^2_0(u)}\,\rbrack +
\nonumber \\ &+& \frac{|\cos\theta|}{\gamma^2} 
\ln \lbrack{\frac{\lambda_{\theta} + \lambda_c |\cos\theta|}
{\lambda_{ab}(1+|\cos\theta|)}}\rbrack
\rbrace 
\label{eq.ap1}
\end{eqnarray}
where  $V_0$ and its derivative are given by,
\begin{eqnarray}
V_0(u) &=& \frac{1}{2\mu_0\tanh\big(\mu_0\sigma/2\big)} + 2
\sum_{m=1}^{\infty}\sum_{s=1}^{\infty} 
\frac{\cos{(m \chi
s)}}{\sqrt{\mu_0^2+m^2}}\exp{(-\sqrt{\mu_0^2+m^2}\,\sigma s)} +
\nonumber \\ 
&+&  \sum_{m=1}^{\infty}\big( \frac{1}{\sqrt{\mu_0^2+m^2}}-\frac{1}{m}
\big) + \ln{(\mu_0/2)}+ Ec \label{eq.ap2}\\ 
\nonumber \\
-\mu_0^2(u) \frac{\partial V_0(u)}{\partial \mu_0^2(u)} &=& \frac{1}{4\mu_0}
\lbrack \frac{1 }{\tanh{\big(\mu_0\sigma/2\big)}} + \frac{\mu_0\sigma/2}
{\sinh^2{(\mu_0\sigma/2)}} \rbrack + \nonumber \\
 &+& \mu_0^2  \sum_{m=1}^{\infty}\sum_{s=1}^{\infty}
\frac{\cos{(m \chi s)}}{\mu_0^2+m^2}\exp{(-\sqrt{\mu_0^2+m^2}\,\sigma s)}
\lbrack \frac{1}{\sqrt{\mu_0^2+m^2}} + \sigma s\rbrack  +\nonumber \\
&+& \frac{\mu_0^2}{2} \sum_{m=1}^{\infty}\big( \frac{1}{(\mu_0^2+m^2)^{3/2}}
\label{eq.ap3}
\end{eqnarray}
where $Ec = 0.5772 \ldots$ is Euler's constant and $\chi$ is defined as
$\chi = (2\pi L_2/L_1)\cos{\varphi}$.
The functions of $c_1$, $c_2$, $\sigma$ and $\mu_0$ are defined as:
\[ c_1(u) = \frac{\lambda^2_{\theta}}{\lambda^2_c} -
u(1-\gamma^2)\cos^2{\theta}\; ,\]  
\[c_2(u) = 1 - u(1-\gamma^2)\; ,\]
\[\sigma
=2\pi\frac{L_2}{L_1}\sqrt{\frac{c_1(u)}{c_2(u)}}\sin{\varphi}\; ,\]
and 
\[\mu_0 =\frac{L_1}{2\pi\lambda_c\sqrt{c_1(u)}} \;. \]

In the expression for $\varepsilon_{in}$, Eq.\ (\ref {eq.ap1}),
$V_0(0)$ stands for $V_0(u)$, Eq.\ (\ref {eq.ap2}), for $u=0$,
or, in other words with the functions $c_1$, $c_2$, $\sigma$ and
$\mu_0$ in this equation evaluated at $u=0$. 

In our numerical calculations of  $\varepsilon_{in}$
we evaluate $V_0(0)$ by using the following generic approximation for the
summations in eqs.(\ref{eq.ap2}) and (\ref{eq.ap3})
\begin{equation}
\sum_{n=1}^{\infty}g(n)\approx\sum_{n=1}^{M}g(n)+\frac{1}{2}
\int_{M}^{\infty}dx(g(x)+g(x+1)),
\label{eq.ap4}
\end{equation}
where $g(x)$ is an analytic function. By using $M=200$ we obtain
excellent numerical precision with modest computer run times.

The integral in eq.(\ref{eq.ap1}) is calculated using  standard
integration algorithms found in the literature, up to a precision of one
part in $10^8$. We run typical calculations of 10-20,000 Metropolis steps
for each run. The results
obtained are easily reproducible.




\end{document}